# Strain, doping and electronic transport of large area monolayer MoS$_2$ exfoliated on gold and transferred to an insulating substrate

S.E. Panasci[1,2], E. Schilirò[1], G. Greco[1], M. Cannas[3], F. M. Gelardi[3], S. Agnello[3,1,4], F. Roccaforte[1], F. Giannazzo[1*]

[1] CNR-IMM, Strada VIII, 5 95121, Catania, Italy

[2] Department of Physics and Astronomy, University of Catania, Via Santa Sofia 64, 95123 Catania, Italy

[3] Department of Physics and Chemistry Emilio Segrè, University of Palermo, Via Archirafi 36, 90123 Palermo, Italy

[4] ATeN Center, Università degli Studi di Palermo, Viale delle Scienze, Edificio 18, 90128 Palermo, Italy

*E-mail: filippo.giannazzo@imm.cnr.it

**Abstract**

Gold-assisted mechanical exfoliation currently represents a promising method to separate ultra-large (cm-scale) transition metal dichalcogenides (TMDs) monolayers (1L) with excellent electronic and optical properties from the parent van der Waals (vdW) crystals. The strong interaction between Au and chalcogen atoms is the key to achieve this nearly perfect 1L exfoliation yield. On the other hand, it may affect significantly the doping and strain of 1L TMDs in contact with Au. In this paper, we systematically investigated the morphology, strain, doping, and electrical properties of large area 1L MoS$_2$ exfoliated on ultra-flat Au films (0.16-0.21 nm roughness) and finally transferred to an insulating Al$_2$O$_3$ substrate. Raman mapping and correlative analysis of the E' and A$_1$' peaks positions revealed a moderate tensile strain ($\varepsilon \approx 0.2\%$) and p-type doping (n$\approx$-0.25$\times$10$^{13}$ cm$^{-2}$) of 1L MoS$_2$ in contact with Au. Nanoscale resolution current mapping and current-voltage (I-V) measurements by conductive atomic force microscopy (C-AFM) showed direct tunnelling across the 1L MoS$_2$ on Au, with a broad distribution of tunnelling barrier values ($\Phi_B$ from 0.7 to 1.7 eV) consistent with the p-type doping of MoS$_2$. After the final transfer of 1L MoS$_2$ on Al$_2$O$_3$/Si, the strain was converted to compressive (ε≈-0.25%). Furthermore, an n-type doping (n≈0.5×10$^{13}$ cm$^{-2}$) was deduced by Raman mapping and confirmed by electrical measurements of an Al$_2$O$_3$/Si back-gated 1L MoS$_2$ transistor. These results provide a deeper understanding of the Au-assisted exfoliation mechanisms and can contribute to its widespread applications for the realization of novel devices and artificial vdW heterostructures.

## 1. Introduction

Semiconducting transition metal dichalcogenides (TMDs) are a class of two-dimensional (2D) layered materials with the general chemical formula MX$_2$, being M a transition metal (Mo, W,..) and X a chalcogen (S, Se,...), which are characterized by strong (covalent) in plane bonds and weak Van der Waals (vdW) interactions between the layers [1]. In particular, due to its abundance in nature and good stability under ambient conditions, molybdenum disulphide (MoS$_2$) has been the most widely investigated TMD for potential applications in electronics, optoelectronics, photodetection and sensing [2,3,4,5]. In its bulk form, MoS$_2$ shows an indirect band gap of 1.2 eV, whereas the monolayer counterpart exhibits a direct band gap of ~1.8 eV [6,7,8,9,10]. The sizeable bandgap, combined with a low dielectric constant, made MoS$_2$ a potential candidate to replace silicon as channel material in ultra-thin body field effect transistors for next generation CMOS applications [11,12,13]. Furthermore, bandgap tunability of MoS$_2$, obtained by tailoring the number of layers [14], strain [15] or dielectric environment [16], offers many possibilities to realize new concept beyond-CMOS electronic devices [17].

Many of the MoS$_2$ device prototypes demonstrated so far have been fabricated using monolayer or few layer flakes obtained by mechanical exfoliation from bulk molybdenite. In spite of the reported progresses in the large area synthesis of TMDs by scalable deposition techniques (including chemical vapour deposition [18,19], molecular beam epitaxy [20] and pulsed laser deposition [21]), mechanical exfoliation still remains a method of choice for investigating basic physical phenomena and to demonstrate new device concepts, due to the superior quality of the material produced by this approach [22,23].

To overcome the limitations represented by the small (micrometer) size of the exfoliated flakes and the lack of reproducibility in the thickness, appropriate strategies allowing to increase the exfoliated monolayer area have been recently elaborated. In particular, the so-called "gold-assisted" mechanical exfoliation approach showed the possibility of separating large area (cm$^2$) monolayer MoS$_2$ (1L

MoS$_2$) from a bulk crystal stamp by exploiting the strong affinity between a gold film and the topmost sulphur atoms of MoS$_2$ [24,25,26,27]. The Au/1L MoS$_2$ stack can be also transferred to insulating substrates and, after Au removal by chemical etching, the large area MoS$_2$ film exhibits electronic properties fully comparable with those of the semiconducting MoS$_2$ flakes obtained by the conventional Scotch tape exfoliation [25,28]. The gold-assisted exfoliation approach has been shown to be effective also with other common TMDs (such as MoSe$_2$, WS$_2$, WSe$_2$, MoTe$_2$, WTe$_2$, and GaSe) [24,26,29] and it has been recently proposed as a general approach to produce large area heterostructures of different TMDs with outstanding electronic quality by sequentially stacking the exfoliated monolayers [29].

In the last few years, several morphological and spectroscopic investigations have been reported on the Au/MoS$_2$ system, with the aim of deeply understanding the mechanisms of the Au-assisted exfoliation and to maximize the monolayer fraction and the lateral size of the obtained MoS$_2$ films. In particular, the 1L exfoliation yield was shown to be strongly influenced by the gold surface morphology and its exposure to the air before exfoliation [24]. Due to the strong vdW interaction at MoS$_2$/Au interface, the Au morphology may significantly affect also the doping and strain in 1L MoS$_2$, as shown by Raman analyses [27]. An increase of the density of states (DOS) at the Fermi energy (i.e. a metallic character) was predicted by ab-initio simulations of the MoS$_2$/Au heterostructure as compared to semiconducting free-standing MoS$_2$ [24]. Such increased DOS in 1L MoS$_2$ associated to the underlying Au was also demonstrated by electrochemical characterization of the MoS$_2$/Au system [24].

In this context, a systematic study on the evolution of the structural and electronic properties of 1L MoS$_2$ in the different stages of the Au-assisted exfoliation process, i.e. after adhesion with gold and after final transfer to an insulating substrate, would be highly desirable.

In this work, we investigated the morphology, strain, doping, and electrical properties of 1L MoS$_2$ exfoliated on ultra-flat Au films and finally transferred to an Al$_2$O$_3$/Si substrate. The correlative analysis of the E' and A$_1$' Raman peaks positions in spatial mapping revealed a moderate tensile strain (~0.2%) and p-type doping (0.25×10$^{13}$ cm$^{-2}$) of 1L MoS$_2$ in contact with Au. Nanoscale resolution current mapping and current-voltage (I-V) measurements by conductive atomic force microscopy (C-AFM) showed direct tunnelling across the 1L MoS$_2$ on Au, with a broad distribution of tunnelling barrier values ($\Phi_B$ from 0.7 to 1.7 eV) indicating wide point to point variations of MoS$_2$ p-type doping. After the final transfer of 1L MoS$_2$ on Al$_2$O$_3$/Si and complete removal of the Au film, the strain was converted to compressive (-0.25%) and a n-type doping of ~0.5×10$^{13}$ cm$^{-2}$ was observed both by Raman and confirmed by electrical measurements on a Al$_2$O$_3$/Si back-gated 1L MoS$_2$ transistor.

## 2. Results and discussion

The lateral extension and thickness uniformity of MoS$_2$ monolayers exfoliated on a smooth gold surface was initially assessed. To this aim, a 15 nm thick Au film was deposited onto a SiO$_2$/Si substrate by DC magnetron sputtering (as schematically reported in Fig.S1). Prior to Au deposition, a 10 nm thick Ni film was sputtered to improve the adhesion onto the SiO$_2$. Beside ensuring an optimal adhesion to the SiO$_2$ surface, the Ni interlayer was beneficial to achieve a very smooth surface of the Au overlayer, with a low root-mean-square (RMS) roughness of 0.16 nm, as deduced from the tapping mode atomic force microscopy (AFM) image reported in Fig.S2 of Supporting Information. Mechanical exfoliation of MoS$_2$ was carried out on the fresh Au surface, i.e. immediately after the deposition, in order to avoid its contamination with adventitious carbon, which is known to reduce the interaction strength between S atoms and Au [24]. By this procedure, very large area MoS$_2$ films, mostly composed of monolayer, were separated from the bulk crystal.

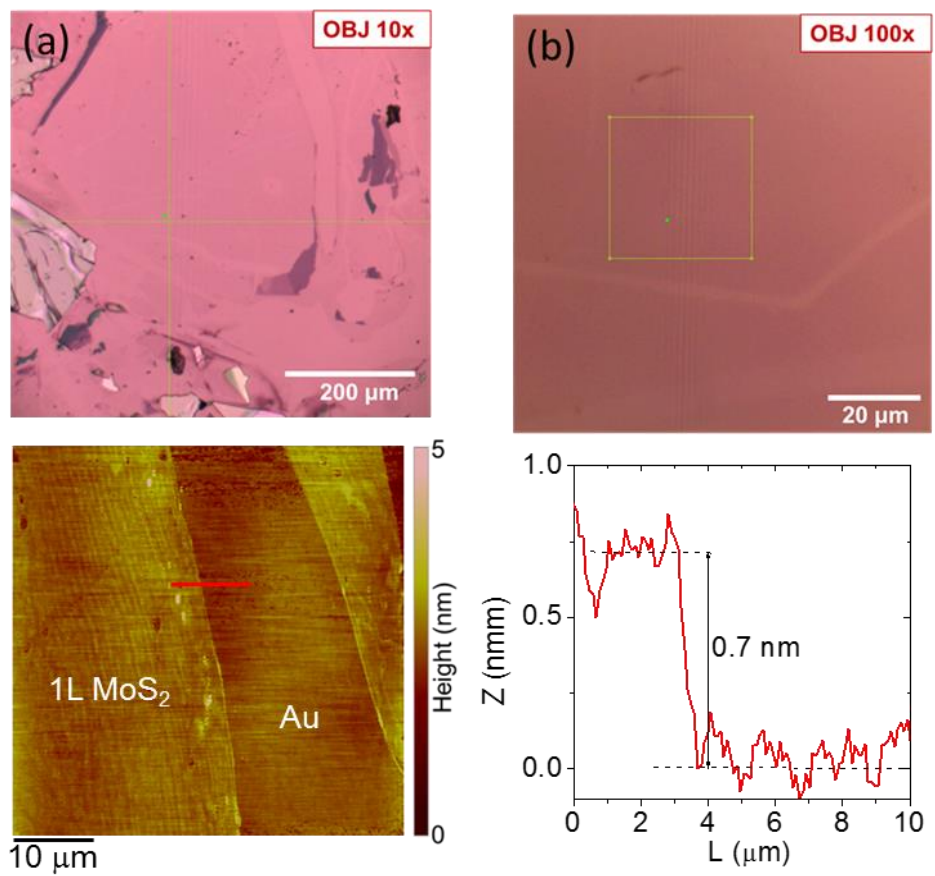

**Fig.1** (a,b) Optical images at two different magnifications, 10× and 100×, of the exfoliated MoS$_2$ on the Au/Ni/SiO$_2$. (c) AFM image of the ultra-thin MoS$_2$ film with a fracture and (d) height profile along the dashed red line. The step height of ~0.7 nm demonstrates the 1L thickness of MoS$_2$.

Fig.1(a) and (b) show two optical images at different magnifications (10× and 100×, respectively) of the exfoliated MoS$_2$ on the Au surface. The presence of an ultra-thin MoS$_2$ film extending for several hundreds of micrometres can be deduced from the colour contrast in the lower magnification image (Fig.1(a)), which also shows the presence of thicker MoS$_2$ areas with smaller size, and of fractures of the MoS$_2$ membrane (i.e. bare Au regions) due to the exfoliation process. The optical contrast difference between the uniform ultra-thin MoS$_2$ membrane and one of these fractures can be better visualized in the higher magnification image in Fig.1(b). Furthermore, a typical tapping mode AFM image of a fracture of the MoS$_2$ film is reported in Fig.1(c). The ~0.7 nm step height measured by the line profile in Fig.1(d) is a direct confirmation of the 1L thickness of the MoS$_2$ membrane.

After assessing the thickness uniformity of 1L MoS$_2$ films exfoliated on gold, we investigated the transfer of these films to an insulating substrate, which is a mandatory requirement for most of electronic applications. More specifically, a Si substrate covered by 100 nm Al$_2$O$_3$ film was employed in this experiment, although the transfer procedure can be easily extended to other semiconductors or dielectric materials. Following the approach recently demonstrated by *Liu et al.* [29], the transfer procedure consisted of three different steps, schematically illustrated in Fig.2. The first step was the fabrication of an ultra-flat "gold tape", consisting of a gold film on a polymer substrate. To this aim, a ~100 nm thick Au layer was deposited by DC magnetron sputtering on an accurately pre-cleaned silicon sample. Afterwards, the Au surface was spin-coated by a protective PMMA layer and attached to a thermal release tape (TRT). By exploiting the poor adhesion between Au and Si, the TRT/PMMA/Au stack was easily peeled from the silicon surface, thus obtaining the desired "gold tape". The surface of Au films prepared by this method is typically very flat [30,31] and it has been already demonstrated to be suitable for the exfoliation of large area monolayers of MoS$_2$ and other TMDs [29]. In particular a RMS roughness of 0.21 nm was evaluated by AFM on the peeled Au films on PMMA in our experiments (see Fig.S3 of Supporting Information), which is comparable with that of the Au/Ni film on SiO$_2$.

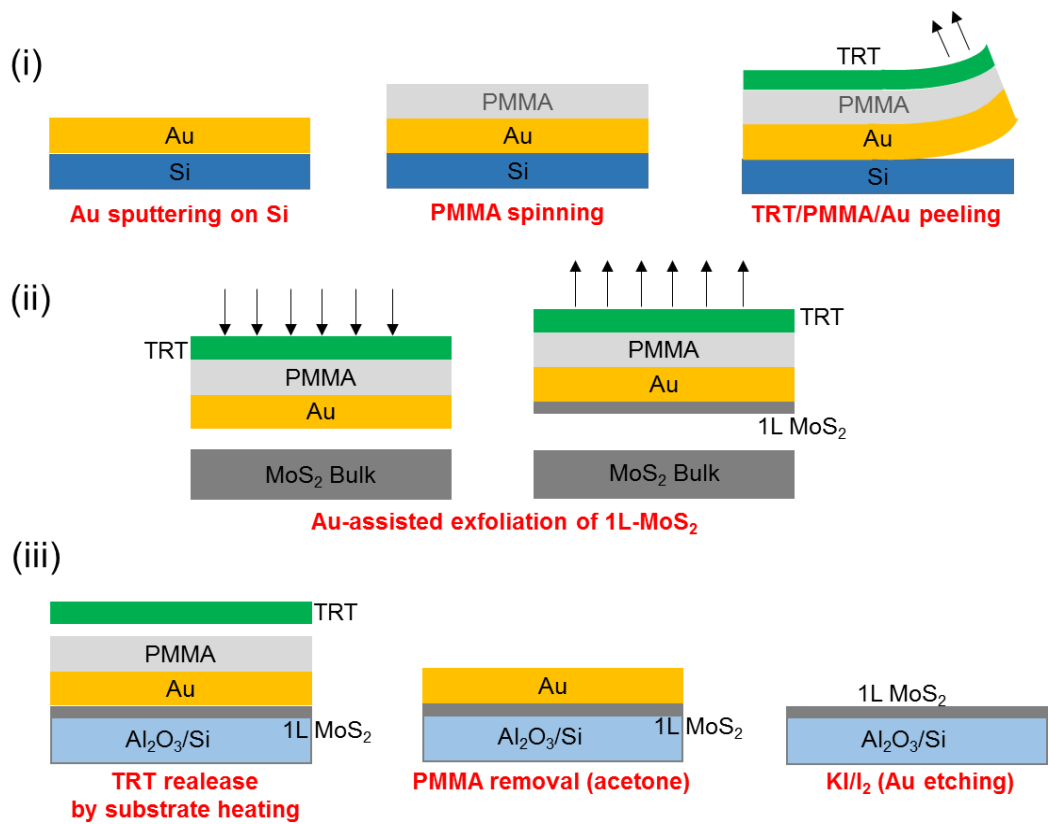

**Fig.2** Schematic illustration of the three steps for Au-assisted exfoliation of 1L MoS$_2$ and transfer to a Al$_2$O$_3$/Si substrate: (i) preparation of a gold tape on PMMA; (ii) 1L MoS2 exfoliation from bulk; (iii) transfer to the final substrate.

The TRT/PMMA/Au stamp with a fresh Au surface, i.e. immediately after peeling from Si, was used to exfoliate 1L MoS$_2$ from a MoS$_2$ bulk sample. The final step of the process was the transfer of 1L MoS$_2$ on the target Al$_2$O$_3$/Si surface. This was achieved by pressing the TRT/PMMA/Au/1L MoS$_2$ stack onto the Al$_2$O$_3$/Si substrate while heating at 120 °C to promote the TRT release, followed by PMMA removal and final chemical etching of the Au film (with KI/I$_2$ solution).

Fig.3(a) reports a typical optical microscopy image of the transferred MoS$_2$ membrane on the Al$_2$O$_3$ surface. As compared to the case of 1L MoS$_2$ exfoliated on gold (Fig.1(a) and (b)), a much sharper colour contrast can be observed between the regions coated by the extended 1L MoS$_2$ membrane (blue) and bare Al$_2$O$_3$ regions (violet), due to the favourable optical interference with the 100 nm Al$_2$O$_3$/Si substrate. Furthermore, the small size regions coated by few-layer or multilayer MoS$_2$ can be easily identified by the azure or bright colour, respectively. Hence, the optical image provides useful information on the thickness uniformity of the transferred MoS$_2$ film on large area. Furthermore, a morphological AFM image of a sample region partially covered by the MoS$_2$ membrane is shown in Fig.3(b). The monolayer MoS$_2$ thickness is directly confirmed by the ~0.7 nm step height measured by the linescan in Fig.3(c).

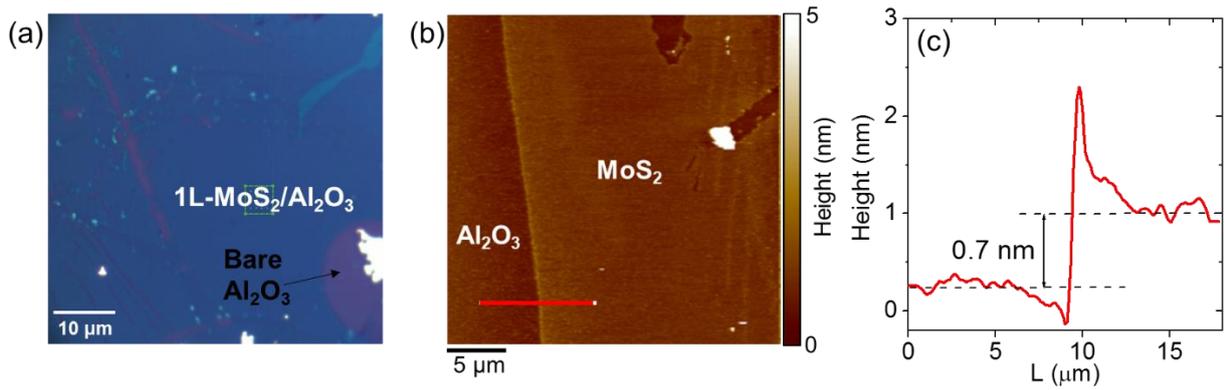

**Fig.3** (a) Optical image (50 μm×50 μm) and (b) AFM morphology on large area (25 μm×25 μm) of the transferred 1L MoS$_2$ membrane on the Al$_2$O$_3$/Si substrate. (c) Height line-scan confirming the monolayer thickness.

The large area 1L MoS$_2$ membranes exfoliated on Au and transferred onto Al$_2$O$_3$/Si were extensively investigated by micro-Raman mapping and photoluminescence (PL) spectroscopy, in order to evaluate the impact of the different substrates on relevant parameters, such as the doping and strain distribution.

Fig.4(a) shows the comparison between two representative Raman spectra for 1L MoS$_2$ on Au (black line) and on Al$_2$O$_3$/Si (red line), with indicated the characteristic E' and A$_1$' peaks, associated to the in-plane and out-of-plane MoS$_2$ vibrational modes, respectively. Noteworthy, a peaks frequency difference $\Delta\omega=18$ cm$^{-1}$ is measured for our large-area 1L MoS$_2$ produced by Au-assisted exfoliation and transferred onto Al$_2$O$_3$, a value very similar to those reported for mechanically exfoliated or CVD grown 1L MoS$_2$ on common insulating substrates (such as SiO$_2$) [23,32]. On the other hand, for the Au-supported 1L MoS$_2$, the E' peak exhibits a red-shift and the A$_1$' peak a blue-shift, resulting in a significantly larger $\Delta\omega=21$ cm$^{-1}$. It is well known that E' and A$_1$' spectral features are highly sensitive to strain and doping of 1L MoS$_2$ [33,34]. In particular, a red shift of E' peak is typically observed with increasing the tensile strain [35,36], followed by a peak splitting for large strain values [27,37]. On the other hand, the A$_1$' peak is known to be sensitive to doping, and a blue (red) shift of its position is typically reported for p-type (n-type) doping of 1L MoS$_2$ [38]. Hence, the increase of $\Delta\omega$ for 1L MoS$_2$ on Au/Ni/SiO$_2$/Si can be ascribed to a change both in the strain and doping of the 2D membrane.

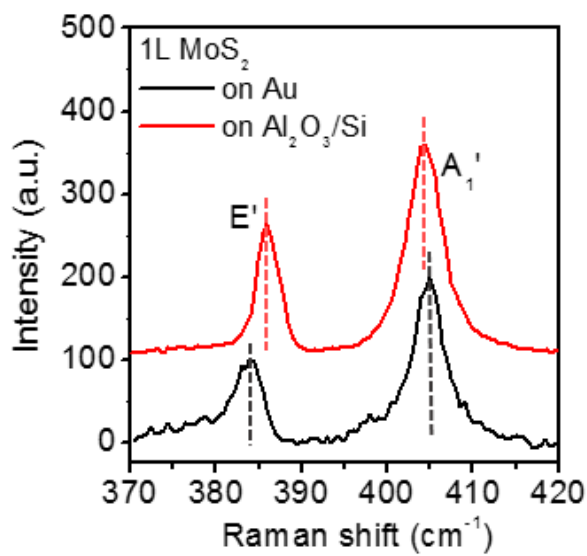

**Fig.4** (a) Representative Raman spectra for 1L MoS$_2$ on Au (black line) and on Al$_2$O$_3$/Si (red line).

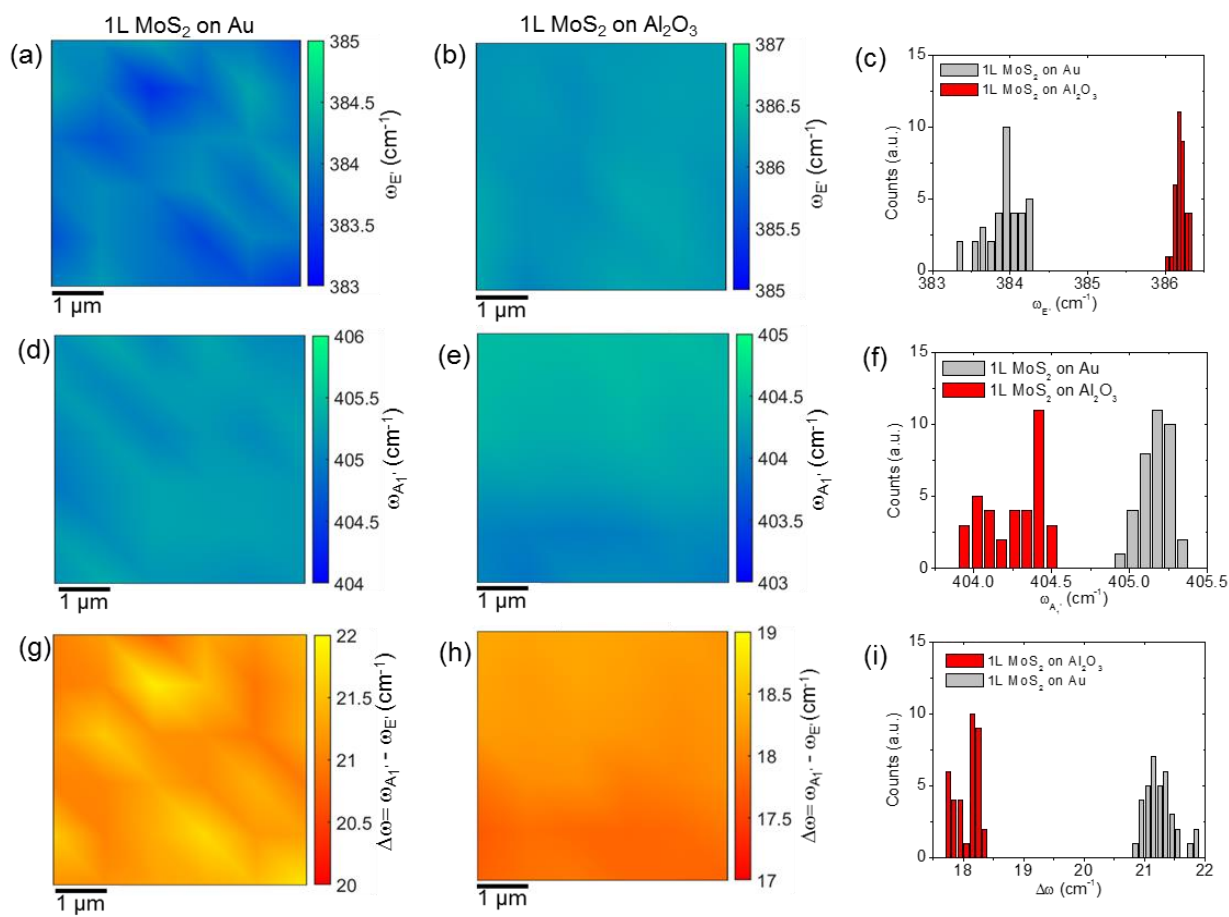

**Fig.5** Colour map of the E' peak frequency values ($\omega_{E'}$) for 1L MoS$_2$ on Au (a) and on Al$_2$O$_3$ (b) and corresponding histograms (c). Colour map of the A$_1$' peak frequency values ($\omega_{A_1'}$) for 1L MoS$_2$ on Au (d) and on Al$_2$O$_3$ (e) and corresponding histograms (f). Colour map of the peaks frequency difference ($\Delta\omega=\omega_{A_1'}-\omega_{E'}$) for 1L MoS$_2$ on Au (g) and on Al$_2$O$_3$ (h) and corresponding histograms (i).

In order to extract relevant statistical information on the doping and strain uniformity of the 1L MoS$_2$ membranes exfoliated on Au and transferred to the Al$_2$O$_3$/Si substrate, Raman mapping was carried out on both samples by collecting arrays of 6×6 spectra on a 5×5 µm$^2$ area. Fig.5(a) and (b) show the colour maps of the E' peak frequency ($\omega_{E'}$) in the scanned areas for 1L MoS$_2$ on Au and Al$_2$O$_3$, respectively, while the comparison between the histograms of the $\omega_{E'}$ values in the two maps is shown in Fig.5(c). Similarly, the colour maps of the A$_1$' peak frequency ($\omega_{A_1'}$) and corresponding histograms are reported in Fig.5(d), (e) and (f). Beside the individual peak positions, also their difference $\Delta\omega=\omega_{A_1'}-\omega_{E'}$ was calculated for all the collected Raman spectra. The colour maps of the $\Delta\omega$ values for 1L MoS$_2$ on Au and Al$_2$O$_3$ are shown in Fig.5(g)-(h), and the histograms of the $\Delta\omega$ values are reported in Fig.5(i).

The comparison between the colour maps allows to visualize the spatial distribution of the $\omega_{E'}$, $\omega_{A_1'}$ and $\Delta\omega$ spectral features in the two different samples. As an example, it can be clearly deduced that the maxima of $\Delta\omega$ for the Au-supported 1L MoS$_2$ sample (Fig.5(g)) are correlated to the minima of the $\omega_{E'}$ map (Fig.5(a)), where the E' peak is more red-shifted. On the other hand, for the 1L MoS$_2$ on Al$_2$O$_3$, the $\Delta\omega$ map exhibits an almost uniform contrast, and the spatial variations are clearly correlated with those of the A$_1$' peak.

The histograms in Fig.5(c) and (f) confirm on a large set of data the red-shift of the E' peak and the blue-shift of the A$_1$' peak for 1L MoS$_2$ on Au with respect to 1L MoS$_2$ on Al$_2$O$_3$. It is also interesting to observe a significantly narrower spread of E' values for the 1L MoS$_2$ transferred to Al$_2$O$_3$, which can be ascribed to a more uniform strain distribution. By Gaussian fitting of the histograms, the average values and standard deviations of the peak frequencies and their difference have been obtained and reported in Table I

|  | $\omega_{E'}$ (cm$^{-1}$) | $\omega_{A_1'}$ (cm$^{-1}$) | $\Delta\omega$ (cm$^{-1}$) | $\varepsilon$(%) | n (10$^{13}$ cm$^{-2}$) |
|---|---|---|---|---|---|
| **1L MoS$_2$ on Au** | 383.9±0.3 | 405.1±0.1 | 21.2±0.3 | 0.21±0.06 | -0.25±0.06 |
| **1L MoS$_2$ on Al$_2$O$_3$** | 386.2±0.1 | 404.2±0.1 | 18.1±0.2 | -0.25±0.01 | 0.5±0.09 |

**Table I** Average values and standard deviation of the E', A$_1$' peaks frequencies ($\omega_{E'}$ and $\omega_{A_1'}$) and their difference ($\Delta\omega$), and of the evaluated strain and doping for 1L MoS$_2$ on Au and on Al$_2$O$_3$.

In the following, the spatial distribution of strain $\varepsilon$ (%) and doping n (cm$^{-2}$) for 1L MoS$_2$ on Au and on Al$_2$O$_3$ will be quantitatively evaluated from a correlative plot of the A$_1$' vs E' peak frequencies for all the Raman spectra in the maps of Fig.5. A similar approach, based on the correlative plot of the characteristic 2D and G peaks, has been widely employed for strain and doping quantification of monolayer graphene on different substrates [39,40,41,42,43]. More recently such method has been adopted by some authors also for 1L MoS$_2$ [33,34,44].

In Fig.6(a), the black open circles represent the A$_1$' vs E' pairs for all the Raman spectra collected on 1L MoS$_2$ on Au, while the blue open triangles represent the data pairs for 1L MoS$_2$ on Al$_2$O$_3$. The red and black lines represent the theoretical relations between the frequencies of the two vibrational modes at a laser wavelength of 532 nm in the ideal cases of a purely strained (strain line) and of a purely doped (doping line) 1L MoS$_2$ [36,38]. The strain and doping lines cross in a point, corresponding to the $\omega_{E'}^0$ and $\omega_{A1'}^0$ frequencies for ideally unstrained and undoped 1L MoS$_2$. In the following, the literature values of the peak frequencies for a suspended MoS$_2$ membrane ($\omega_{E'}^0 = 385\ cm^{-1}$ and $\omega_{A1'}^0 = 405\ cm^{-1}$) [36] have been kept as the best approximation to these ideal values, as substrate effects are excluded in this case. Starting from this reference point, the directions of increasing tensile strain and n-type doping are also indicated by the arrows along the two lines.

The ε and n values for each experimental point in Fig.6(a) can be evaluated from the combination of the linear relationships between the biaxial strain/charge doping and Raman shifts of the vibrational modes:

$$\omega_{E'} = \omega_{E'}^0 - 2\gamma_{E'}\omega_{E'}^0 \varepsilon + k_{E'} n \qquad (1a)$$

$$\omega_{A1'} = \omega_{A1'}^0 - 2\gamma_{A1'}\omega_{A1'}^0 \varepsilon + k_{A1'} n \qquad (1b)$$

Here, $\gamma_{E'}$=0.68 and $\gamma_{A1'}$=0.21 are the Grüneisen parameters for the two vibrational modes of 1L MoS$_2$ [36,45,46]. The $k_{E'}$ = -0.33×10$^{-13}$ cm and $k_{A1'}$ = -2.2×10$^{-13}$ cm coefficients are the shift rates of Raman peaks as a function of the electron concentration n (in cm$^{-2}$) in 1L MoS$_2$, obtained by Raman characterization of electrochemically top-gated MoS$_2$ transistors [38].

In particular, the relation for the strain line can be obtained by solving the system of Eqs (1a)-(1b) in the case of n=0:

$$\omega_{A1'} = \omega_{A1'}^0 + \frac{\gamma_{A1'}\omega_{A1'}^0}{\gamma_{E'}\omega_{E'}^0}\left(\omega_{E'} - \omega_{E'}^0\right) \qquad (2)$$

whereas the doping line equation is obtained by the same procedure for ε=0:

$$\omega_{A1'} = \omega_{A1'}^0 + \frac{k_{A1'}}{k_{E'}}\left(\omega_{E'} - \omega_{E'}^0\right) \qquad (3)$$

Hence, $\frac{\gamma_{A1'}\omega^0_{A1'}}{\gamma_{E'}\omega^0_{E'}} = 0.32$ and $\frac{k_{A1'}}{k_{E'}} = 6.67$ are the slopes for the strain and doping lines, respectively.

The dashed red lines parallel to the strain line (n=0) and the dashed black lines parallel to the doping line (ε=0) serve as guides to quantify the doping and strain values, respectively. They correspond to ±0.1% variations for the strain and ±0.1×10$^{13}$ cm$^{-2}$ variations for the doping. Since the ω$_{E'}$ is more sensitive to biaxial strain [36], the spacing between the dashed black lines parallel to the doping line is calculated from the E' mode strain rate $2\gamma_{E'}\omega^0_{E'}$ =5.2 cm$^{-1}$/% . On the other hand, since the A$_1$' mode results mainly influenced by charge doping [36], the spacing between the dashed red lines parallel to the strain line is calculated from the A$_1$' doping rate $k_{A1'}$=0.22×10$^{-13}$ cm.

The plot in Fig.6(a) shows that all the experimental data points for 1L MoS$_2$ on Au are located above the strain line and in the left side with respect to the doping line. Hence, as compared to the reference case of a free-standing (suspended) 1L MoS$_2$, our gold-supported 1L MoS$_2$ films exhibit a tensile strain in the range from ~0.1% to ~0.3% and a p-type doping in the range from ~0.1×10$^{13}$ to ~0.4×10$^{13}$ cm$^{-2}$. The average values of the strain (~0.21%) and doping (~0.25×10$^{13}$ cm$^{-2}$) are indicated by the black square in Fig.6(a). A tensile biaxial strain, originating from the lattice mismatch between MoS$_2$ and Au [47,48], has been recently observed in the case of 1L MoS$_2$ exfoliated on Au also by other authors [27], that reported very large ε values up to 1.2%. The smaller tensile strain obtained in our samples are probably due to the very smooth surface of the gold films. The observed p-type doping of MoS$_2$ in contact with Au is consistent with several recent reports of a p-type behavior induced by MoS$_2$ functionalization with gold nanoparticles, adsorbates or Au-based chemicals [49,50,51].

On the other hand, the cloud of data for 1L MoS$_2$ on Al$_2$O$_3$ is located in a region of the ε-n plane corresponding to a compressively strained and n-type doped film, with the strain values comprised in a narrow range around ~-0.25% and the electron density ranging from ~0.4×10$^{13}$ to ~0.7×10$^{13}$ cm$^{-2}$. The compressive strain can be plausibly related to the transfer procedure and the adhesion properties of 1L-MoS$_2$ with the Al$_2$O$_3$ surface. The observed n-type doping is consistent with the unintentional doping typically observed for MoS$_2$ layer on insulating substrates, and can be ascribed, in part, to charge transfer by adsorbed or interface trapped charges under ambient conditions, as well as to native defects of MoS$_2$.

By solving Eq.(1a) and (1b) for all the data points of the ω$_{E'}$ and ω$_{A1'}$ maps, the corresponding colour maps of the strain (Fig.6(b)-(c)) and doping (Fig.6(e)-(f)) for the two samples were obtained. The corresponding histograms of the strain and doping values are reported in Fig.6(d) and (g), respectively. From the comparison of the strain and doping maps on the 1L MoS$_2$/Au, a correlation between the regions with higher tensile strain and those with higher p-type doping can be noticed. This suggests that both strain and p-type doping originate from a locally stronger interaction with Au. On the other hand, the compressive strain distribution appears very uniform in the 1L MoS$_2$ membrane transferred onto Al$_2$O$_3$, without any clear correlation with the doping distribution. The average values and standard deviation of the strain and doping for the two different samples have been extracted by Gaussian fitting of the histograms in Fig.6(d) and (g), and the obtained values have been reported in the Table I. Obviously, the spatial resolution in these maps is limited by the laser spot size (~1 µm). Furthermore, the concentration sensitivity (in the order of $10^{12}$ cm$^{-2}$) is limited by the shift rate of the $A_1'$ peak with doping concentration. Higher spatial resolution and sensitivity information on the doping distribution in the Au supported 1L MoS$_2$ will be deduced from conductive atomic force microscopy analyses reported later on in this paper.

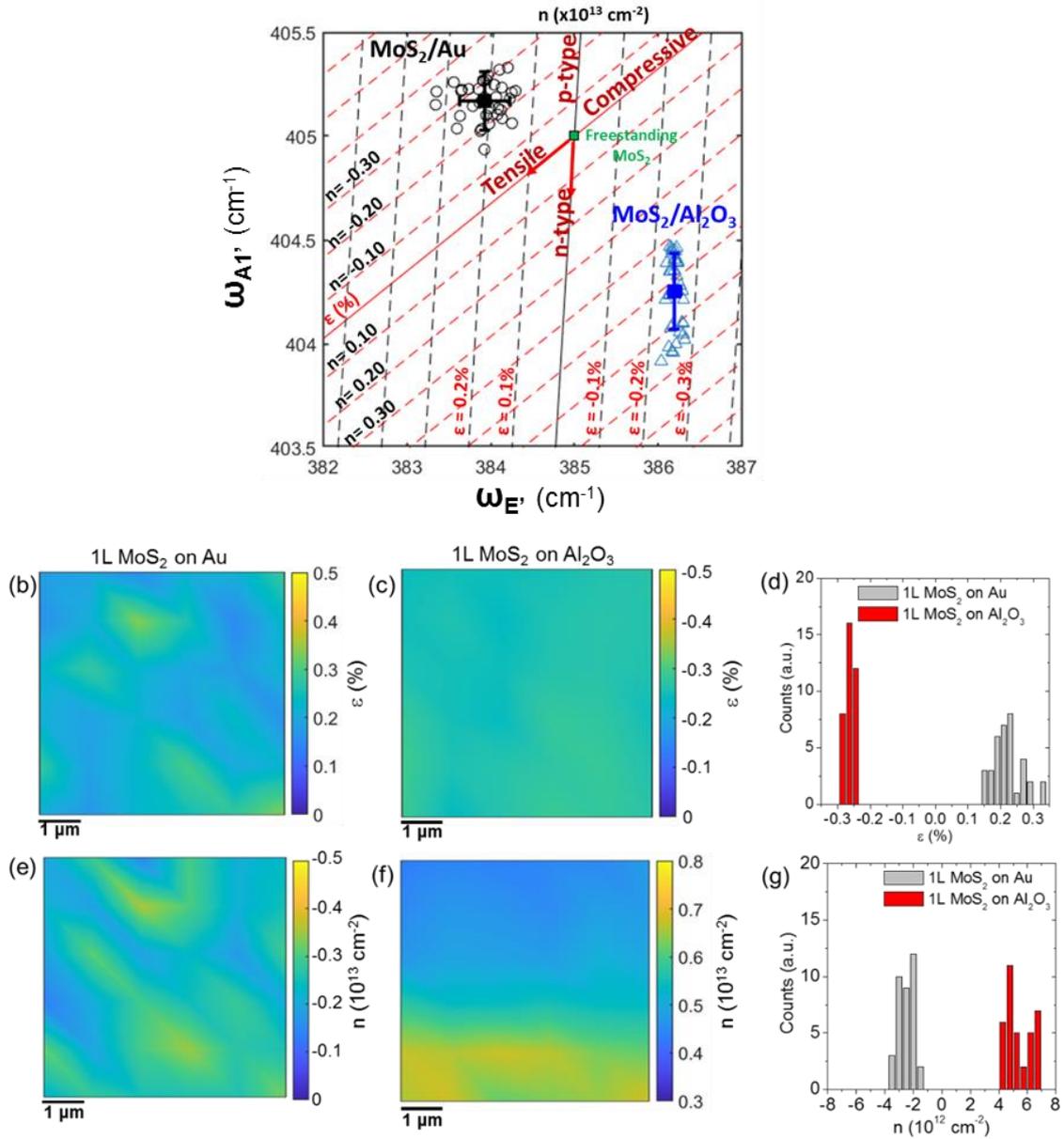

**Fig.6**. (a) Correlative plot of the $A_1'$ and $E'$ peak frequencies to evaluate the biaxial strain and charge doping distributions in 1L $MoS_2$ on Au (black circles) and on $Al_2O_3$ (blue triangles). The red (black) lines represent the strain (doping) lines for ideally undoped (unstrained) 1L $MoS_2$, meanwhile the green square indicates the $\omega_{E'}^0 = 385\ cm^{-1}$ and $\omega_{A_1'}^0 = 405\ cm^{-1}$ frequencies for freestanding 1L $MoS_2$, taken as zero reference. The dashed red (black) lines parallel to the strain (doping) lines serve as guides to quantify the doping and strain values, respectively. Colour maps of the strain for the 1L $MoS_2$ on Au (b) and 1L $MoS_2$ on $Al_2O_3$ (c) samples and histograms of the strain values (d). Colour maps of the doping for 1L $MoS_2$ on Au (e) and 1L $MoS_2$ on $Al_2O_3$ (f) and histograms of the doping values (g).

To further investigate the impact of the substrate/$MoS_2$ interaction on the electronic properties of 1L $MoS_2$, micro-photoluminescence analyses were also performed using the 532 nm laser probe of the Raman equipment as excitation source. Fig.7 shows the comparison between two representative PL spectra collected on the two different samples under the same illumination conditions. Noteworthy, the large 1L MoS$_2$ membrane produced by gold-assisted exfoliation and finally transferred onto Al$_2$O$_3$ exhibits a prominent peak at 1.83 eV, very similar to that observed in monolayer MoS$_2$ obtained by the traditional mechanical exfoliation or deposited by CVD. On the other hand, a strongly reduced PL intensity was observed when the exfoliated 1L MoS$_2$ membrane is still in contact with Au, together with a red-shift of the main PL peak to 1.79 eV. Such red shift has been also confirmed by comparison of arrays of PL spectra collected on the two samples, as reported in the colour maps in Fig.S4 of Supporting Information. The strong reduction of the PL intensity is consistent with the emission quenching reported by other authors for 1L MoS$_2$ exfoliated on Au [24] and for MoS$_2$ functionalized with Au nanoparticles [52]. This PL quenching can be explained in terms of a preferential transfer of photo-excited charges from MoS$_2$ to Au. In addition, the tensile strain of the MoS$_2$ layer in contact with Au can also play a role in the reduction of the PL yield, and the observed ~40 meV red shift of the main PL peak is consistent with the measured strain variation of $\Delta\varepsilon \approx 0.4\%$ between the two samples [36].

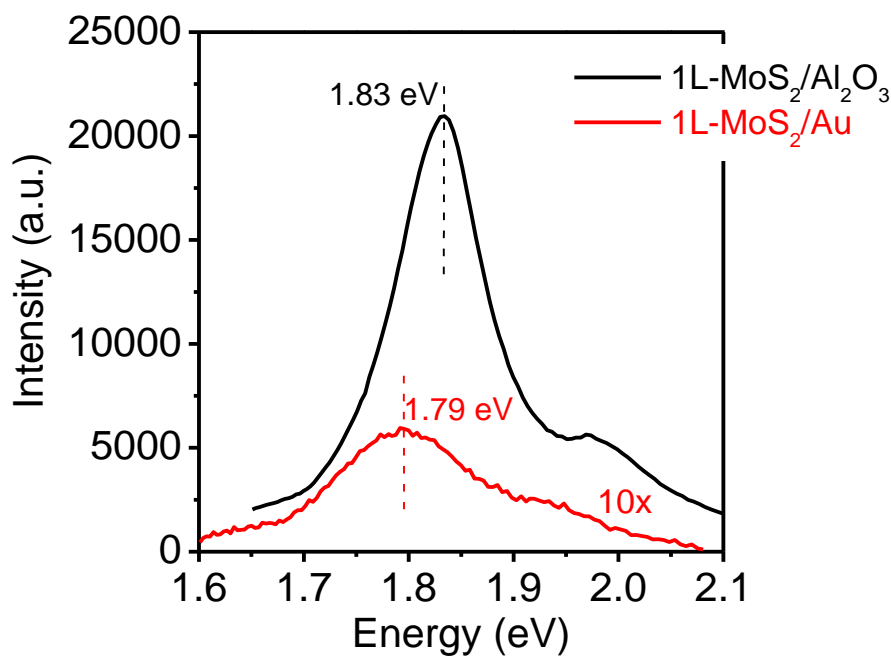

**Fig.7**. Typical microPL spectra collected under excitation at 532 nm on 1L MoS$_2$ on Au (with the intensity multiplied by a factor of 10) and 1L MoS$_2$ transferred to Al$_2$O$_3$.

Then, a nanoscale resolution electrical characterization of the Au-supported 1L MoS$_2$ membrane by C-AFM measurements [53] was carried out to get further information on the doping uniformity in this ultra-thin layer. To this aim, the current injection at the interface between the Au substrate and 1L MoS$_2$ was probed at nanoscale by a Pt-coated Si tip, according to the configuration schematically illustrated in Fig.8(a). The surface morphology in a sample region partially covered by 1L MoS$_2$ is reported in Fig.8(b). The distribution of height values extracted from this map (Fig.8(c)) shows two components, associated to the bare Au surface and 1L MoS$_2$ on Au, respectively, that were fitted by Gaussian peaks with nearly identical full width at-half-maximum. This confirms how the 1L MoS$_2$ membrane conformally follows the smooth Au morphology. Furthermore, Fig.8(c) shows the simultaneously measured current map, collected applying a DC bias $V_{tip}$=50 mV between the Pt tip and the Au electrode (substrate). For this low bias value, the current level measured on the bare Au region reaches the current amplifier saturation limit, whereas appreciable lateral variations of the injected current through the 1L MoS$_2$ membrane can be observed.

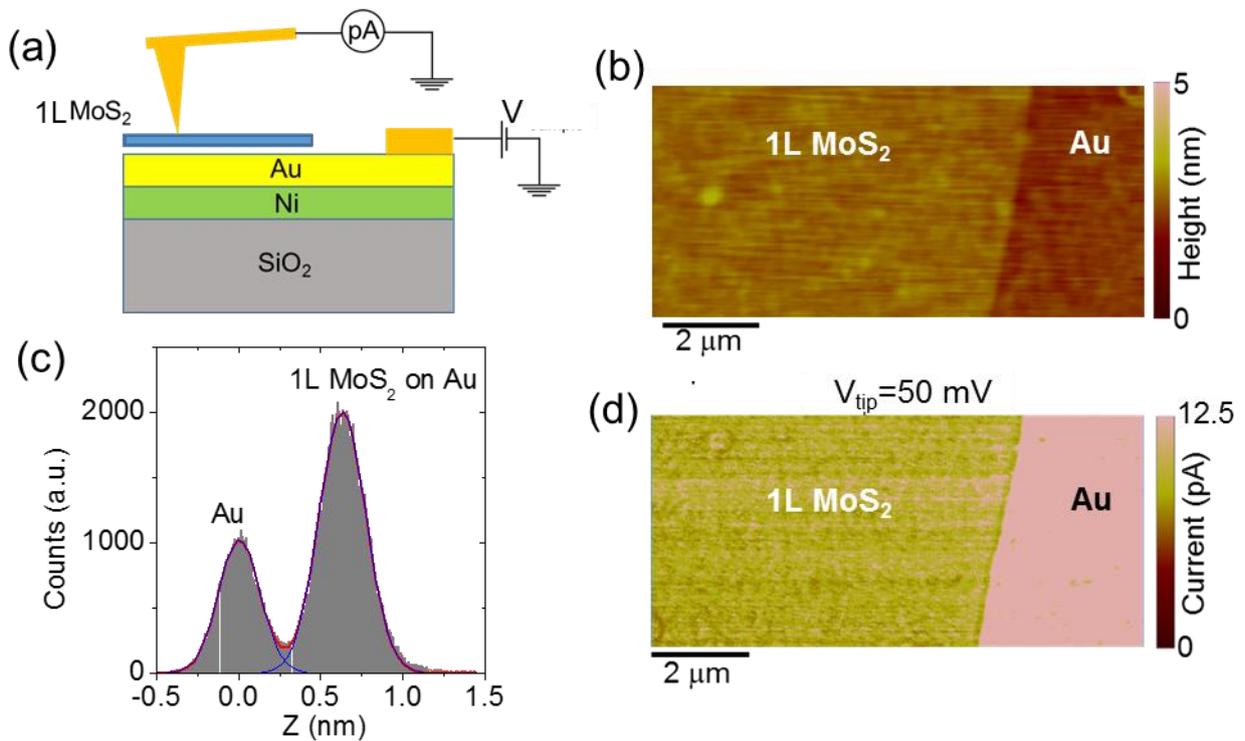

**Fig.8** (a) Schematic illustration of the C-AFM setup used for current mapping through the 1L MoS$_2$ film on Au. (b) Morphology and (c) histogram of height distribution on a sample region with the Au substrate partially covered by the 1L MoS$_2$ film. (d) Simultaneously measured current map on the same area (at $V_{tip}$=50 mV).

Such local variations of the injected current through the atomically thin membrane can be ascribed to the lateral inhomogeneities of electronic properties. To further investigate this aspect and the current transport mechanisms, a set of local current-voltage (I-$V_{tip}$) characteristics where acquired both on the bare Au surface and at different positions on the MoS$_2$ film, as reported in Fig.9(a).

I-$V_{tip}$ curves measured by the Pt tip in contact with Au (see red curve in Fig.9(a)) are very reproducible and exhibit an Ohmic behaviour with a very steep slope and current saturation at few mV positive and negative bias (as shown in the insert of Fig.9(a)). On the other hand, the curves measured on MoS$_2$ show a linear behaviour around $V_{tip}$=0 V with significant variations in the slope at different positions. The slight asymmetry of the I-V$_{tip}$ curves between positive and negative polarizations at larger bias can be ascribed to the different workfunctions of Pt and Au metals.

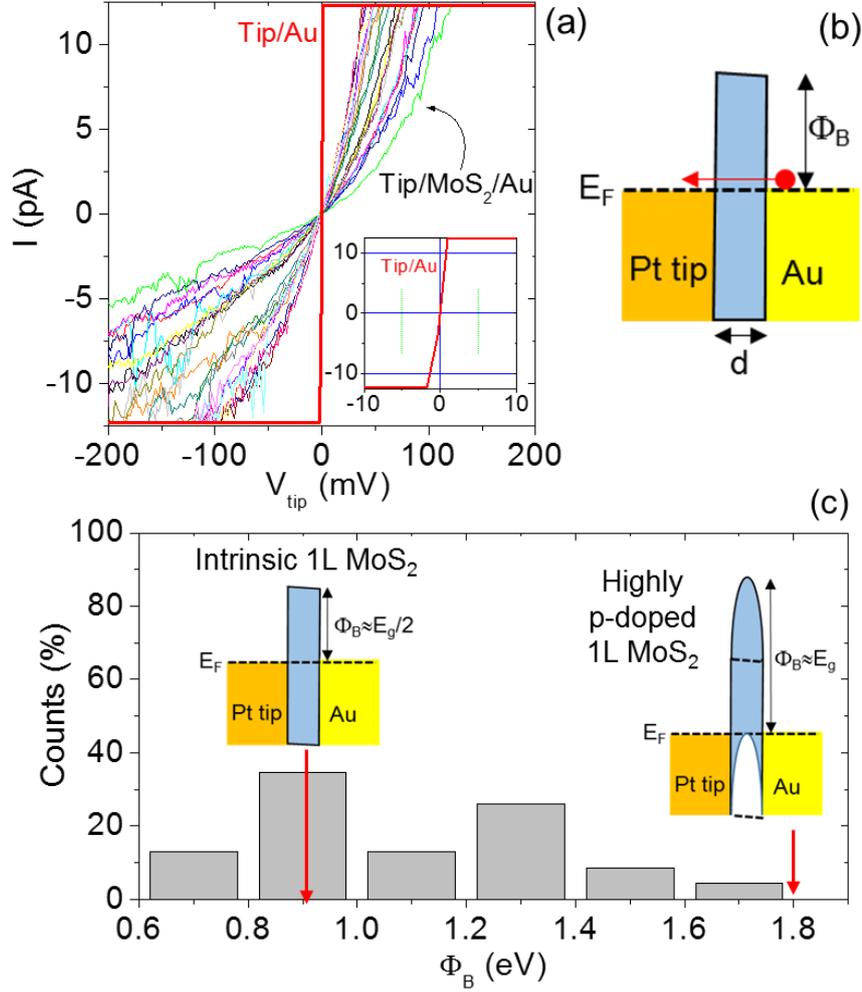

**Fig.9** (a) Local I-V$_{tip}$ curves measured with the Pt-tip in contact with 1L MoS$_2$ on Au and with the bare Au surface (red line). A detail of the I-V$_{tip}$ curve measured on Au is reported in the inset. (b) Schematic band-diagram for the tip/1L MoS$_2$/Au metal/semiconductor/metal heterojunction. (c) Histogram of the tunnelling barrier values Φ$_B$ evaluated from the I-V$_{tip}$ curves in panel (a), according to the direct tunnelling mechanism. The band-diagram for intrinsic and highly p-type doped 1L MoS$_2$ are schematically illustrated in the inserts of (c).

The system formed by the tip in contact with the 1L MoS$_2$/Au can be described as a metal/semiconductor/metal heterojunction. The linear behaviour of the I-V$_{tip}$ characteristics indicates direct tunnelling (DT) as the most appropriate mechanism ruling current transport in this heterostructure. In particular, the tunnelling current can be expressed as:

$$I_{DT} = BV_{tip} \times P(\Phi_B, d) = BV_{tip}\, exp\left[-\frac{4\pi\sqrt{2m_{eff}\Phi_B}\,d}{h}\right] \qquad (4)$$

where B is a pre-factor (proportional to the tip contact area) and $P(\Phi_B,d)$ is the direct tunnelling probability, which is a function of the tunnelling barrier thickness d (i.e. the MoS$_2$ thickness) and its height $\Phi_B$, corresponding to the energy difference between the MoS$_2$ conduction band and the Au Fermi level (see schematic in Fig.9(b)). Here m$_{eff}$=0.35 m$_0$ is the electron effective mass in the transversal direction for 1L MoS$_2$ [54] and h is the Planck's constant. As a matter of fact, the thickness dependent tunnelling probability becomes unity when the MoS$_2$ layer is absent (d=0), i.e. when the tip is directly in contact with the Au substrate. Since current mapping and local I-V measurements have been performed using the same tip in a sample area including MoS$_2$-covered and uncovered Au regions, the same value for the pre-factor B were considered in the two cases. Hence, the experimental values of the local tunnelling probability at different positions on MoS$_2$ were estimated as the ratio between the slope of the I-V curves measured on MoS$_2$ and the slope of the I-V characteristics measured on Au. Since the MoS$_2$ layer is very conformal to the smooth Au morphology, we have assumed a laterally uniform 1L MoS$_2$ barrier thickness d=0.65 nm (corresponding to the ideal value for 1L MoS$_2$) over the C-AFM probed area. As a result, the local barrier height values have been extracted from the tunnelling probabilities for each of the I-V curves in Fig.9(a). The obtained histogram of the $\Phi_B$ values, reported in Fig.9(c), shows a broad distribution, extending from ~0.70±0.08 to ~1.70±0.08 eV. In particular, the component at $\Phi_B$≈0.9 eV in this distribution corresponds to a Fermi level located approximately at E$_g$/2 with respect to the MoS$_2$ conduction band, as schematically illustrated in the left insert of Fig.9(c). Noteworthy, this value is very close to ideal barrier height between Au and undoped MoS$_2$, given by $\Phi_B$=W-χ, being W≈5.1 eV the gold work function and χ≈4.2 eV the electron affinity of 1L MoS$_2$ [55]. On the other hand, the tail at larger $\Phi_B$ values in the distribution of Fig.9(c) can be ascribed to the local p-type doping (i.e., the upward bending of the valence and conduction bands) of 1L-MoS$_2$ induced by the Au substrate. In particular, in the regions with the highest p-type doping, the barrier height approaches the value of the bandgap ($\Phi_B$≈E$_g$), as schematically illustrated in the right insert of Fig.9(c). Hence, the C-AFM characterization reveals variations in the local p-type doping of the Au-supported 1L MoS$_2$ membrane over a broad range.

Finally, the electronic transport in 1L MoS$_2$ membrane transferred onto the Al$_2$O$_3$ dielectric surface has been investigated by electrical characterization of a field effect transistor (FET) with the Al$_2$O$_3$ (100 nm)/Si back-gate and Au source and drain contacts (channel length L=10 μm), as illustrated in the insert of Fig.10(a). The output characteristics (drain current vs drain bias, I$_D$-V$_D$) of the device for different gate bias values ranging from V$_G$=-20 to 10 V are shown in Fig.10(a). At low drain bias (V$_D$<3V) current injection in the MoS$_2$ channel is limited by the high Schottky barrier at Au/MoS$_2$ contacts, whereas a linear behaviour of the I$_D$-V$_D$ characteristics is observed at intermediate V$_D$ values

(from 3 to 10 V), followed by current saturation at higher voltages. The transfer characteristic ($I_D$-$V_G$) at a drain bias $V_D=5V$ (i.e. in the linear region of $I_D$-$V_D$ curves) is reported in Fig.10(b), black line. The monotonic increase of $I_D$ with $V_G$ is the typically observed behaviour for a transistor with an n-type $MoS_2$ channel. A negative threshold voltage $V_{th} \approx -8$ V was evaluated by linear fitting of the $I_D$-$V_G$ curve and taking the intercept with the voltage axis, as indicated by the arrow in Fig.10(b). Since $V_{th}$ represents the bias necessary to deplete the n-type $MoS_2$ channel, the electron density can be estimated as $n=C_{ox}|V_{th}|/q$, where $C_{ox}=\varepsilon_0\varepsilon_{ox}/t$ is the $Al_2O_3$ capacitance per unit area, with $\varepsilon_0$ the vacuum permittivity and $\varepsilon_{ox}=8$ is the relative dielectric constant of the $Al_2O_3$ dielectric. The obtained carrier density $n \approx 3.1 \times 10^{12}$ cm$^{-2}$ is in reasonably good agreement with the carrier density values obtained by Raman mapping.

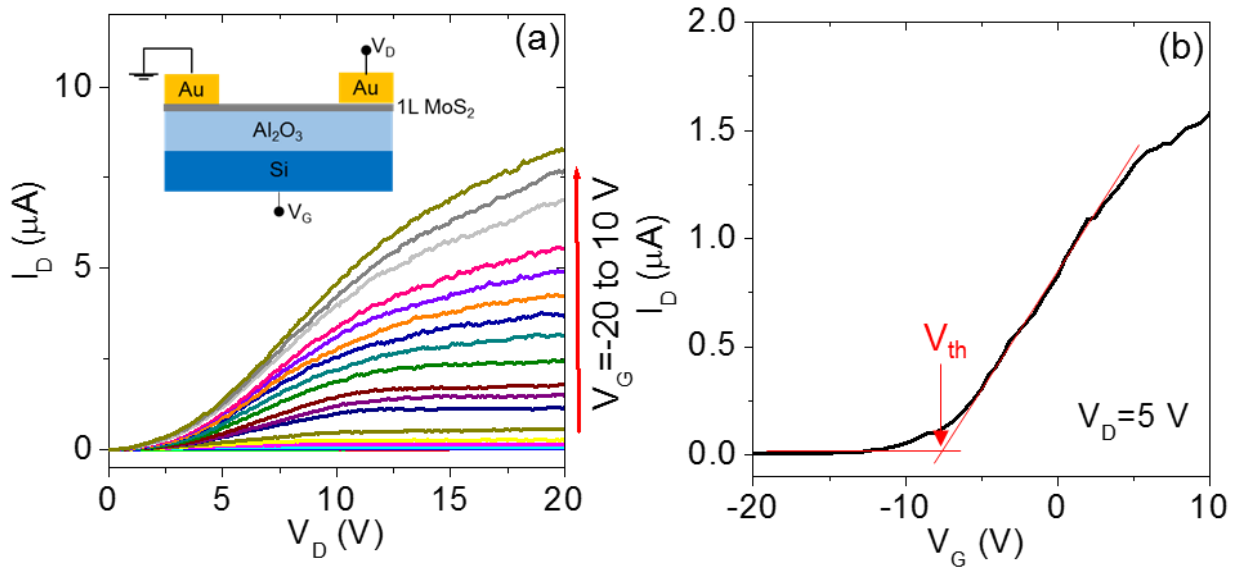

**Fig.10** (a) Output and (b) transfer characteristics of a back-gated field effect transistor fabricated with Au-exfoliated 1L $MoS_2$ transferred on $Al_2O_3$/Si. The device schematic is shown in the insert of panel (a).

Hence, also this device-level characterization confirmed that the large area 1L $MoS_2$ films produced by Au-assisted exfoliation, after the final transfer to an insulating substrate and complete removal of gold, recover the same electrical properties of the $MoS_2$ flakes produced by the conventional mechanical exfoliation or by chemical vapour deposition.

## Conclusion

In summary, large area (cm$^2$) 1L MoS$_2$ membranes have been exfoliated on very flat gold films and transferred to an insulating Al$_2$O$_3$/Si substrate. For 1L MoS$_2$ in contact with Au, Raman mapping revealed a spatially inhomogeneous distribution of tensile strain (in the range from ~0.1% to ~0.3%) and p-type doping (from ~0.1×10$^{13}$ to ~0.4×10$^{13}$ cm$^{-2}$), with a correlation between regions showing higher strain and doping. The electrical properties of Au-supported MoS$_2$ were probed at nanoscale by C-AFM, showing direct tunnelling across the ultra-thin 1L-MoS$_2$, with a broad distribution of tunnelling barrier values ($\Phi_B$ from 0.7 to 1.7 eV) consistent with an inhomogeneous p-type doping of MoS$_2$. After the final transfer of 1L MoS$_2$ on Al$_2$O$_3$/Si, the strain was converted to compressive ($\varepsilon \approx -0.25\%$) with a very uniform distribution. Furthermore, an n-type doping (n≈0.5×10$^{13}$ cm$^{-2}$) was deduced by Raman and confirmed by electrical measurements of an Al$_2$O$_3$/Si back-gated 1L MoS$_2$ transistor. These results provide a deeper understanding of the properties of large area 1L MoS$_2$ produced by Au-assisted exfoliation, and will contribute to the widespread application of this outstanding quality material in the demonstration of novel device concept and synthetic Van der Waals heterostructures.

## Materials and Methods

**Samples preparation.** The deposition of Ni(10 nm)/Au (15 nm) on the SiO$_2$(900 nm)/Si sample was carried out by DC magnetron sputtering using a Quorum equipment. The base vacuum in the chamber was ~10$^{-5}$ mbar, meanwhile during the deposition process the pressure was of about 10$^{-4}$-10$^{-3}$ mbar. The same equipment was employed to deposit 100 nm Au on a Si sample for the preparation of the gold tape with the peeling technique (see Fig.2). PMMA (200K, 0.5 µm) was spin coated on Au and tempered at 150°C. A Nitto Denko thermal release tape (with 120 °C release temperature) was used for the handling of the PMMA/Au gold tape. The 100 nm Al$_2$O$_3$ insulator on Si (used as a final substrate for 1L MoS$_2$ transfer) was deposited by DC-pulsed RF reactive sputter.

**AFM and CAFM analyses.** Morphological analyses of the Au/Ni substrates and of the exfoliated MoS2 films were carried out by tapping mode Atomic Force Microscopy (AFM) using a DI3100 equipment by Bruker with Nanoscope V electronics. Sharp silicon tips with a curvature radius of 5 nm were used for these measurements. C-AFM measurements were carried out with the same AFM system equipped with the TUNA module and using Pt coated Si tips.

**Micro-Raman and micro-Photoluminescence.** Raman spectroscopy and PL measurements were carried out using a Horiba HR-Evolution micro-Raman system with a confocal microscope (100× objective) and a laser excitation wavelength of 532 nm. The laser power used for these analyses was filtered with a neutral density filter at 1% ensuring no sample degradation. A grating of 1800 lines/mm was employed to acquire Raman spectra in a range from 150 to 650 cm$^{-1}$, while a grating of 600 lines/mm was used to acquire Photoluminescence spectra in a range from 10 to 5500 cm$^{-1}$. All the spectra were calibrated with respect to the Silicon peak at 520.7 cm$^{-1}$.

**Field effect transistor preparation and characterization:** A back-gated 1L MoS$_2$ field effect transistor was fabricated with the Au-exfoliated film transferred onto Al$_2$O$_3$ (100 nm)/Si by sputtering Au source/drain contacts with a shadow mask. The contact spacing, i.e. the channel length was L=10 µm. The output and transfer characteristics of the transistor were measured in dark conditions using a Cascade Microtech probe station with an Agilent 4156b parameter analyzer.

**Acknowledgements**

The authors acknowledge S. Di Franco (CNR-IMM) for the assistance in the samples preparation, P. Fiorenza and R. Lo Nigro (CNR-IMM) and M. Cannas (Univ. of Palermo) for useful discussions. The paper has been supported, in part, by MUR in the framework of the FlagERA-JTC 2019 project "ETMOS". E.S. acknowledges the PON project EleGaNTe (ARS01_01007) for financial support. Part of the experiments have been carried out using the facilities of the Italian Infrastructure Beyond Nano.